\documentclass[modern]{aastex631}
\usepackage{wrapfig, needspace}
\usepackage{hyperref}
\hypersetup{linkcolor=red,citecolor=blue,filecolor=cyan,urlcolor=magenta}

\begin{document}

\title{Comet 73P/Schwassmann-Wachmann 3's Dust Trail as a Source of Pickup Ions}

\author[0000-0002-3499-8142]{S. R. Grant}
\affiliation{Mullard Space Science Laboratory\\ Department of Space \& Climate Physics\\ University College London\\ Holmbury St. Mary\\ Dorking, Surrey RH5 6NT, UK}
\email{samuel.grant.20@ucl.ac.uk, g.h.jones@ucl.ac.uk, toni.galvin@unh.edu}

\author[0000-0002-5859-1136]{G. H. Jones}
\affiliation{Mullard Space Science Laboratory\\ Department of Space \& Climate Physics\\ University College London\\ Holmbury St. Mary\\ Dorking, Surrey RH5 6NT, UK}
\affiliation{The Centre for Planetary Sciences at UCL/Birkbeck\\ London, UK}

\author[0000-0003-3752-5700]{A. B. Galvin}
\affiliation{University of New Hampshire\\Durham, NH, USA}



\begin{abstract}

73P/Schwassmann-Wachmann 3 is a short-period comet that has undergone multiple fragmentation events in the last few decades. During May-June 2006, while passing near Earth, multiple fragments of comet 73P passed sunward of Sun-Earth Lagrange Point 1, while cometary pickup ions were detected concurrently by instruments on both the ACE and Wind spacecraft, implying the crossing of one or more ion tail. Additionally, during August 2011, a fragment of 73P passed directly sunward of spacecraft STEREO-B. A detection of cometary ions is shown to originate at fragment 73P-AM. Solar wind velocity measurements are used to extrapolate the flow of the solar wind in 3 dimensions and, when compared with the positions of known comets and cometary fragments, estimate the separation between the cometary ion tail and the spacecraft. Using this technique, it is shown that the alignment of the major cometary fragments with the spacecraft was poor for the transport of cometary ions via the solar wind, but the encounter was near enough for immersion in the diffuse ion tail surrounding an extended dust trail within which the nucleus fragments reside. This implies that, at this distance, the extended trail of cometary debris was a significant source of cometary ions in the case of comet 73P. 

\end{abstract}




\section{Introduction} \label{sec:intro}

The material forming the cometary dust tail and ion tail can be intercepted and measured by spacecraft at significant distances from the source nucleus, providing a probe of the cometary nucleus's composition. The ion tail structure in particular can be detected at multiple Astronomical Units (AU) from the cometary nucleus, as with spacecraft Ulysses's encounter with comet Hyakutake on 1\textsuperscript{st} May 1996, identified by \citet{jonesbalogh2000}, and also reported upon by \citet{gloeckler2000}. 

Cometary ion tail encounters can provide scientifically valuable data on the composition of the comet, coma, and the solar wind. Encounters by spacecraft are surprisingly commonplace \citep{jonesbalogh2000,gloeckler2004, neugebauer2007, matteini2021solar, grant2022}, but are often hard to identify based solely on in-situ plasma measurements, especially when occurring unexpectedly. Currently known encounters, such as the 2006 Ulysses-McNaught encounter \citep{neugebauer2007}, are identifiable due to the exceptionally high density of cometary pickup ions, although the source comet may not always be unambiguously identified \citep{gloeckler2004}.  

A few methods to predict cometary ion tail encounters have been developed, by comparing the orbits of the comet and spacecraft or looking for spacecraft crossing the comet's orbital plane \citep{spjuth2005search, jones2020prospects}. In May 2020, the  ESA (European Space Agency) Solar Orbiter spacecraft was predicted by \citet{jones2020prospects} to cross the orbital plane anti-sunward of comet ATLAS (C/2019 Y4) on 6\textsuperscript{th} June 2020. When the data from Solar Orbiter's instruments from around this time were analysed, all was consistent with a crossing through a cometary ion tail \citep{matteini2021solar}. This makes this event the first serendipitous ion tail crossing to be predicted in advance and subsequently confirmed by instrumentation, although direct measurement of cometary ions was not possible, as not all spacecraft experiments were operational at that time. Comet ATLAS was a fragmented comet; early in 2020, its nucleus broke into multiple fragments and sub-fragments \citep{jones2020prospects, hui2020observations}. Subsequently, the method has been applied to predict and confirm an ion tail encounter with comet C/2021 A1 (Leonard) by Solar Orbiter \citep{grant2022}.

\begin{figure}
    \centering
    \includegraphics[width=0.9\columnwidth]{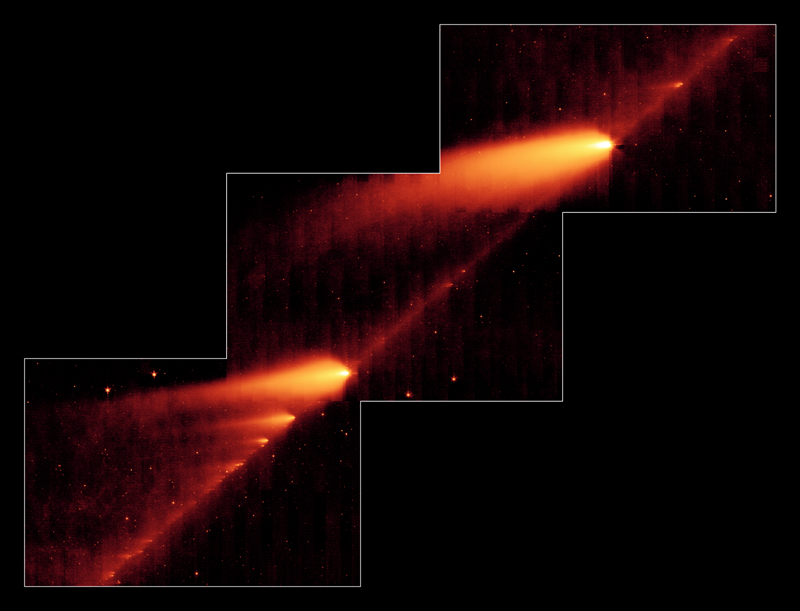}
    \caption{Infrared image from NASA's Spitzer Space Telescope of fragmented comet 73P/Schwassman-Wachmann 3. A trail of material can be seen spanning the space between the large fragments. From \cite{spitzer}.}
    \label{fig:dust_trail_example}
\end{figure}

Cometary fragmentation near perihelion is not rare and can be caused by tidal forces, outgassing, or just gradual weakening of the nucleus. After the initial fragmentation, the resultant fragments  often break apart further, until hundreds of fragments of varying sizes and activity levels are spread along the original comet's orbital path. Slight deviations in trajectory and speed lead the trail and fragments to deviate gradually from the original orbit, and the fragmentation process exposes the chemically pristine cometary interior. Compared to a whole comet, the dust trail can be far more pronounced, at least temporarily, as huge amounts of micro-scale debris are released in the fragmentation process \citep{graykowski2019fragmented, fernandez2009s}.

Comet 73P/Schwassmann-Wachmann 3 is a periodic Jupiter-family comet that has undergone multiple fragmentation events since its discovery in May 1930. Having split into 5 fragments in 1995, further fragmentation occurred during the comet's 2006 apparition, resulting in 70 documented fragments and sub-fragments, with many more undetectable \citep{ho200873p, weaver2008hubble}. In May-June 2006, such fragments passed sunward of Sun-Earth L1, where the ACE and Wind spacecraft were stationed. As shown in Fig. \ref{fig:dust_trail_example}, from May 4 - 6 2006 NASA's Spitzer telescope captured images of the major fragments of comet 73P, with the dust trail and a speckling of deviated fragments visible \citep{spitzer}. A significant flux of cometary pickup ions was recognised in ACE/SWICS (Solar Wind Ion Composition Spectrometer) and Wind/STICS (SupraThermal Ion Composition Spectrometer) data, and was identified as a cometary ion tail crossing by \citet{gilbert2015}. 


Using the method described below, this paper investigates the hypothesis that outgassing from the extended debris trail of comet 73P is the source of the signatures in 2006 identified by \citet{gilbert2015}. A comparable approach in August 2011 between STEREO-B and 73P is presented to provide further context for the encounter in 2006.

\section{Tailcatcher method for the prediction of cometary ion tail immersion}

The Tailcatcher program was developed to quantify the degree of alignment between the ion tail of a comet and a spacecraft. This is done by the calculation of an impact parameter, that is minimised when the alignment is optimised \citep{tailcatcher}. This method provides relatively accurate information on serendipitous spacecraft-comet encounters, using in-situ data to find and refine our understanding of tail and coma crossings. Solar wind velocity measurements made by a spacecraft anti-sunward of a comet can be used to extrapolate the path of the solar wind back toward the Sun. The minimum distance between these extrapolated solar wind packets and known comets is a measure of the likelihood of an ion tail crossing. This method provides a better picture of the geometry of an encounter than do solar wind speed estimates, and can uncover previously unidentified encounters in large data-sets. Crucially, the approximations made during the extrapolation are not significant enough to alter the efficacy of the method. This method is in agreement with observations for all past identified ion tail crossings, and has twice been used to predict future ion tail encounters for two comets with the spacecraft Solar Orbiter \citep{jones2020prospects, grant2022}.

The program does not attempt to directly calculate the broadening of the ion tail, or the size of the coma. However, both of these factors mean that a crossing is also possible to occur in a broader time frame before and after the minimum impact parameter, rather than just at the minimum. Additionally, the method relies on the assumption that the solar wind velocity measured by the spacecraft is the same as that experienced by the comet, and everywhere between the two objects. In reality, the solar wind evolves as it travels, and a solar wind packet will not maintain the same 3-dimensional velocity with radial distance from the Sun. This method is not valid for sun-grazing comets, and demonstrates increased inaccuracies for comet-spacecraft separations of multiple astronomical units. To account for the broadening of the ion tail and the uncertainties in the method, a conservative threshold of 0.05AU is used when judging the likelihood of a tail crossing. This is approximately double the estimated diameter of the ion tail of comet C/1996 B2 (Hyakutake), encountered by Ulysses at a distance of 3.8 AU from the cometary nucleus \citep{jonesbalogh2000}.

The Tailcatcher method is adapted and applied here to investigate unusual cometary pick-up ion encounters of the spacecraft ACE and Wind with the fragments of comet 73P, identified by \citet{gilbert2015}, and a similar encounter of subfragment 73P-AM with spacecraft STEREO-B. 
The JPL Horizons database is the source of all the positional data used in this method \citep{horizons, giorgini1997}.

\section{Observations - ACE/Wind}

\begin{figure}[t]
\centering\includegraphics[width=1.\linewidth]{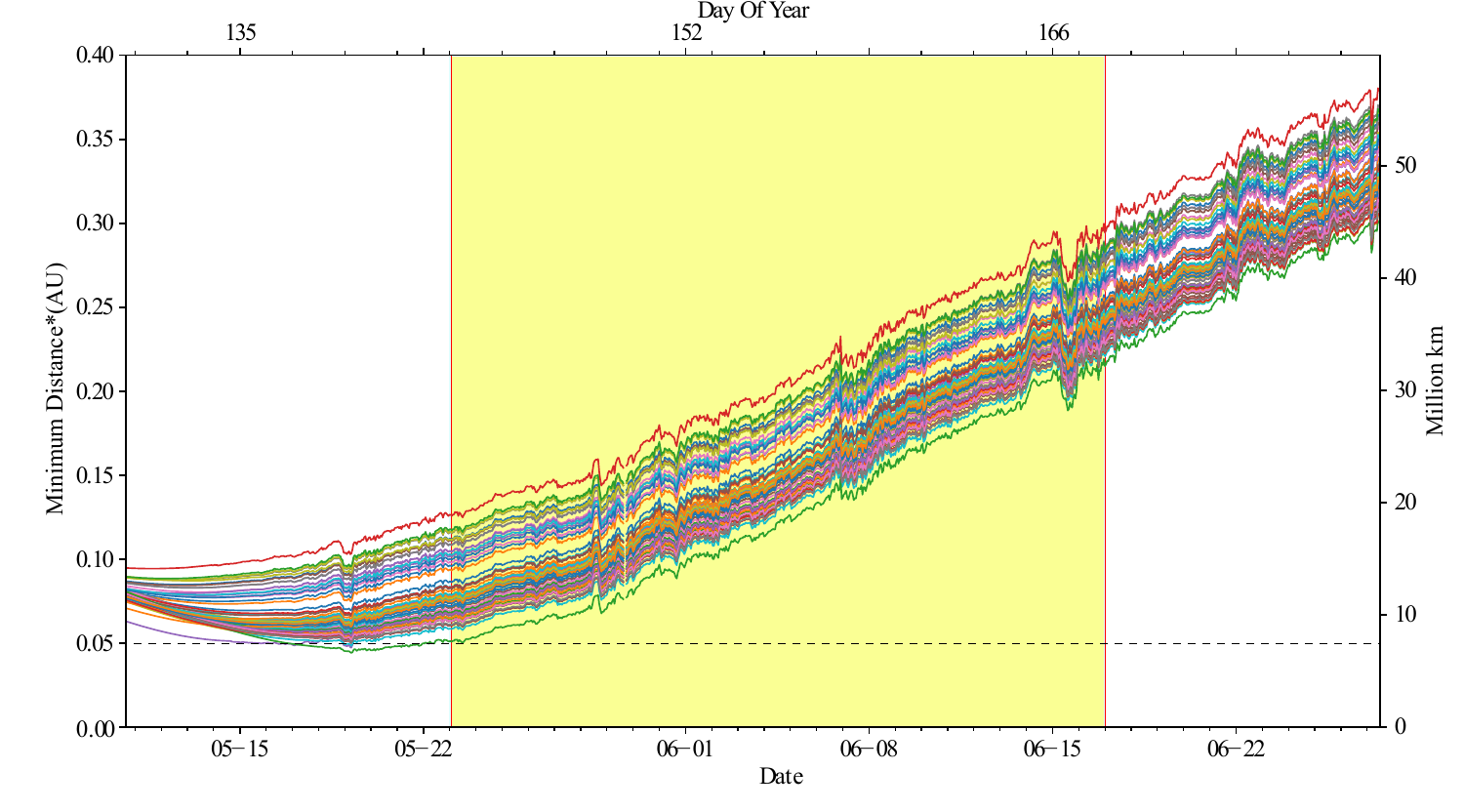}
\caption{73P fragment impact parameters, calculated using the Tailcatcher method, for Sun-Earth Lagrange Point 1, SEL1. Different colours represent the 71 known fragments of 73P at the time of detection. The $x$ axis is the spacecraft time, corresponding to the time at which a propagating SW plasma packet arrives at the spacecraft. The highlighted timeframe corresponds to the measurement of significant O$^{+}$ flux by both ACE and Wind \citep{gilbert2015}. }
 \label{fig:impa2}
\end{figure}

\begin{figure}[ht!]
\centering\includegraphics[width=1.\columnwidth]{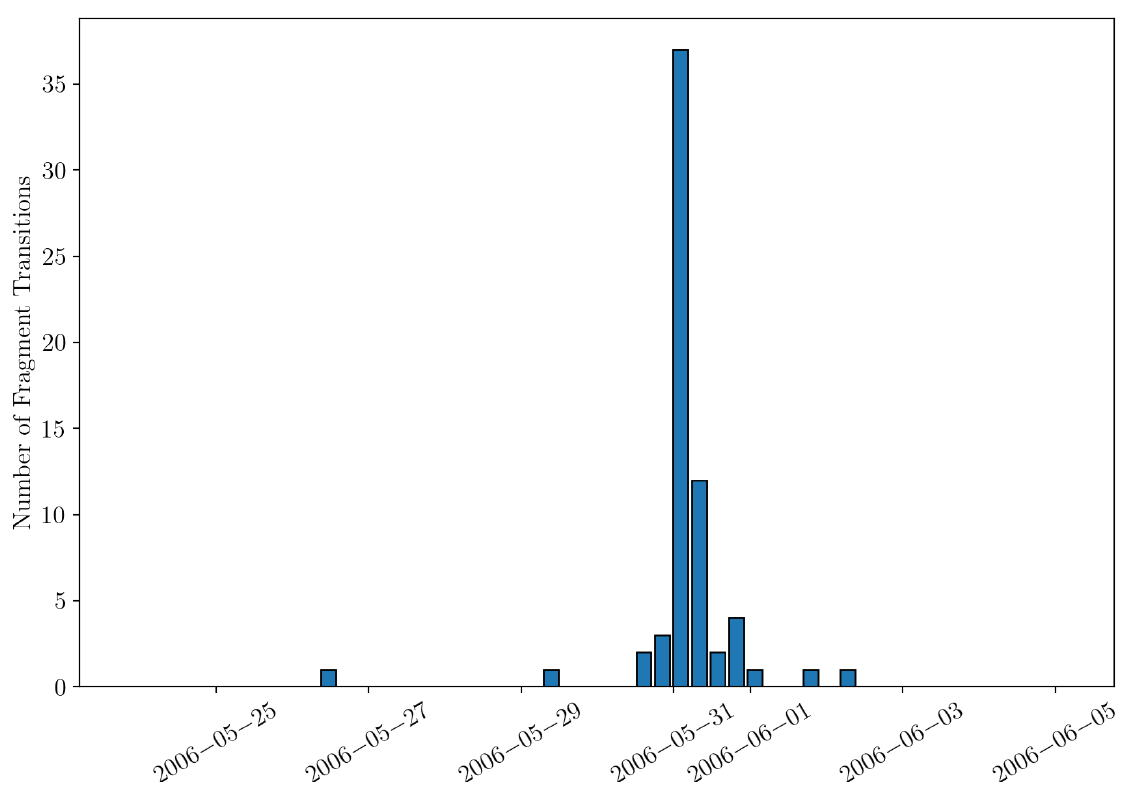}
\caption{Histogram with 6-hour bins showing the binned times at which ACE crosses the orbital planes of the fragments of comet 73P.}
 \label{fig:hist}
\end{figure}

\begin{figure}[ht!]
\centering\includegraphics[width=0.99\linewidth]{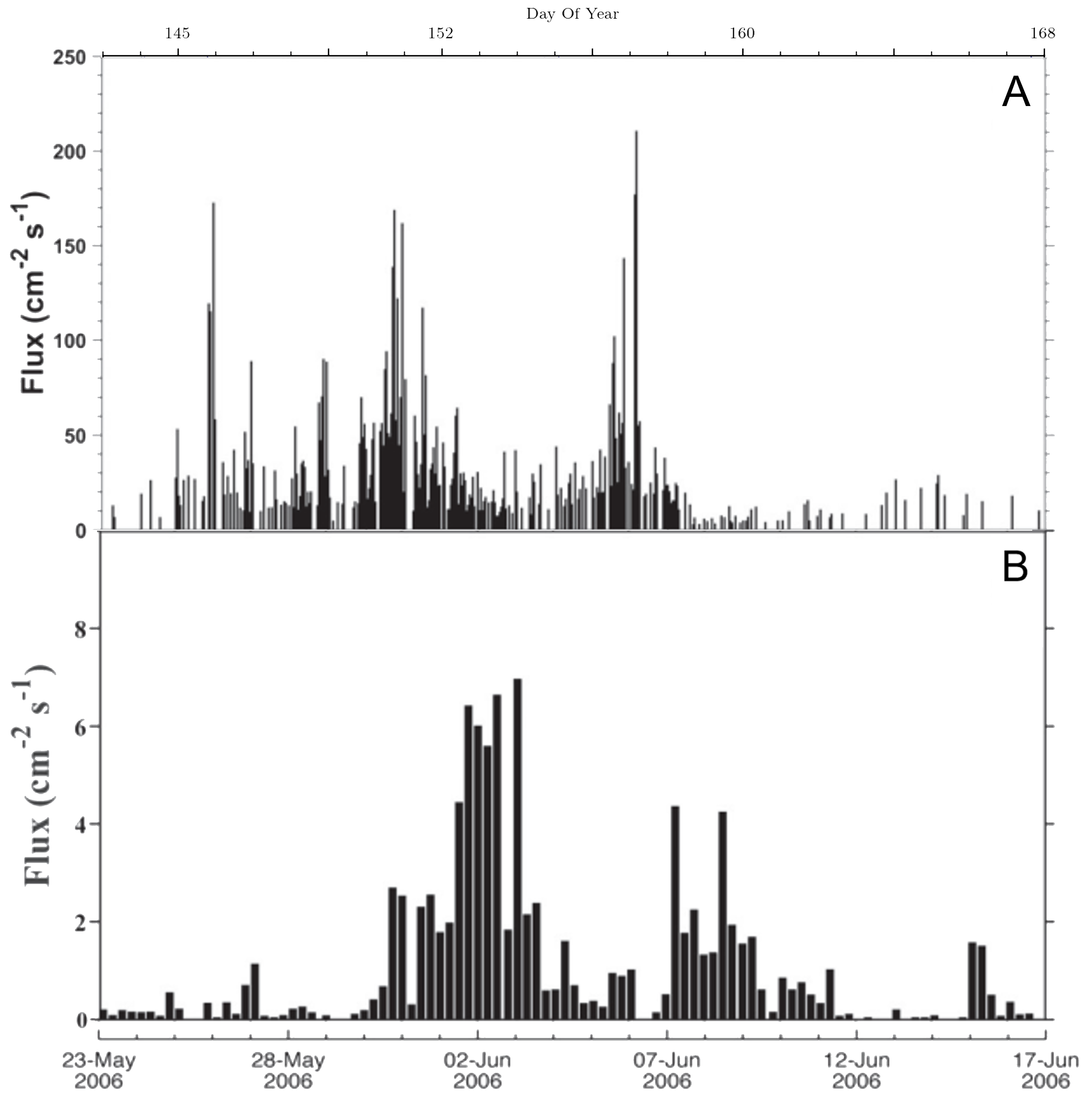}
\caption{Plots A and B show incoming cometary O$^{+}$ flux measured by ACE/SWICS at energies 0.6 - 83 keV $e^{-1}$, and Wind/STICS at energies below 220 keV, respectively. Retrieved from \citet{gilbert2015}. }
 \label{fig:mag_combined}
\end{figure}

\begin{figure}[t!]
    \centering
    \gridline{\fig{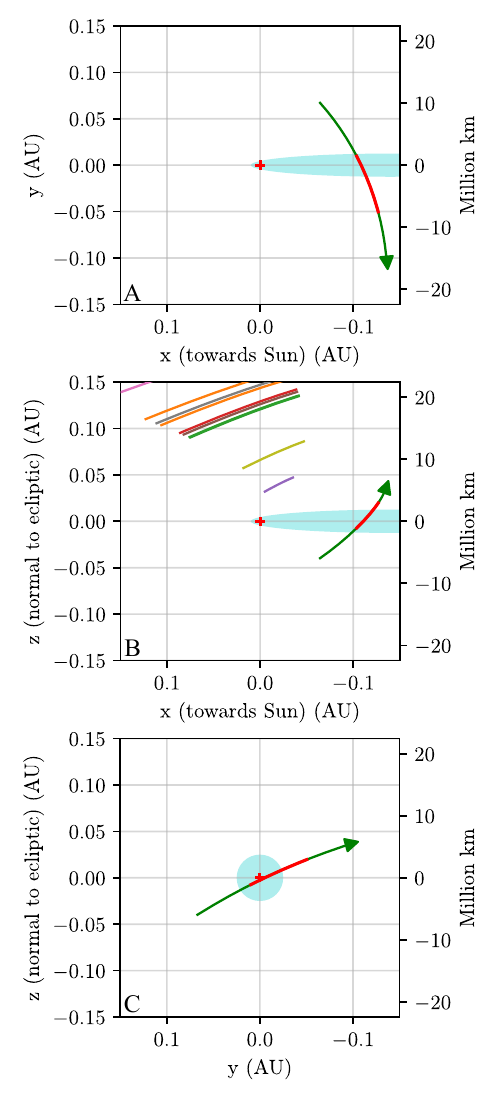}{0.48\columnwidth}{
    \caption{Relative motion of STEREO-B in a frame of reference centered on comet 73P fragment AM. The spacecraft track covers the period in 2011 from 08-29 to 09-18, ending at the triangle. The region where significant O$^{+}$ ion detection (06/09 - 13/09) occurred is highlighted in red. The blue region is a visual aid showing the antisunward direction with respect to the origin. The motions of the other tracked fragments of 73P relative to fragment AM are also plotted. }\label{fig:3dstb}}
    \fig{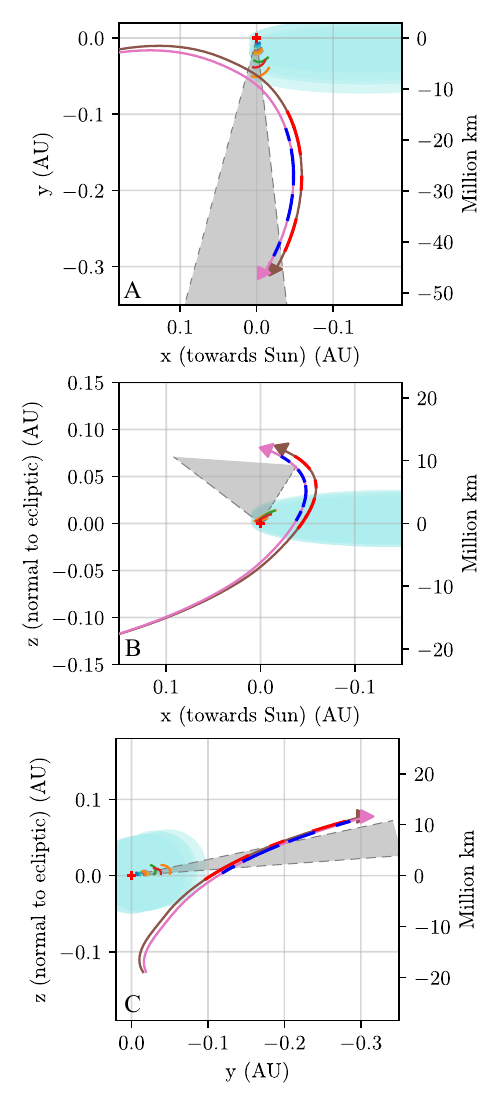}{0.48\columnwidth}{
    \caption{Relative motion of ACE and Wind in a frame of reference centered on comet 73P fragment B. The spacecraft tracks cover the period in 2006 from 04-20 to 06-20, ending at the triangles. The area swept by the extended trail of the fragments is approximated in grey. The prominent regions of significant O$^{+}$ ion flux are highlighted in red and blue for ACE and Wind respectively. For ACE in red, the periods highlighted are 24-05 to 01-06, 05-06 to 07-06 and 12-06 to 17-06. For Wind in blue, the periods highlighted are 26-05 to 28-05, 30-05 to 05-06, 07-06 to 11-06 and 15-06 to 17-06.}\label{fig:3dax}}}

\end{figure}

Fig. \ref{fig:hist} show the binned times when the tracked fragments of comet 73P pass the Sun to Sun-Earth Lagrange point 1 (SEL1) line, with the majority passing on May 31. As identified and reported by J. A. Gilbert et al. in 2015, the notable flux of cometary O$^{+}$ ions by Wind/STICS and ACE/SWICS extends from May 23 until June 17 2006, with four notable maxima \citep{gilbert2015}. The O$^{+}$ ion flux data is plotted in the upper panels of Fig. \ref{fig:mag_combined}. The motion of the two spacecraft during the detection of each flux peak is shown in Fig. \ref{fig:3dax}. 
While the central part of this increase in ion flux, observed from May 23 through June 1, can be attributed to ACE and Wind passing behind the known fragments of 73P, the ion flux increase begins 8 days before the fragments cross the Sun-SEL1 line, and continues for over 2 weeks afterward. This implies some ion source in addition to the large fragments.

The Tailcatcher method is applied here to SEL1 and the fragments of comet 73P. This method approximately accounts for the variable speed and direction of the solar wind flow, so can be used in the 73P –ACE/Wind encounter to determine whether any unusually high tangential solar wind flow can account for the time difference between the ion flux increase and fragment transitions. The impact parameter for each tracked fragment of 73P was calculated using the Tailcatcher method described above, using the 3D proton velocity measurements from each spacecraft, to estimate the flow of the solar wind to spacecraft ACE and Wind. The resulting parameters are plotted in Fig. \ref{fig:impa2}, and compared to the O$^{+}$ flux measurements from ACE/SWICS and Wind/STICS.

The minima, as shown in Fig. \ref{fig:impa2}, do not align well with the timing of the O$^{+}$ flux peaks. This is evidence that the tracked larger fragments are not the sole source of this increased ion flux, and the possibility of an encounter with an ion “sheet” produced by the extended debris trail of the comet is enforced, as discussed later in the text.


Fig. \ref{fig:3dax} shows the trajectories of the spacecraft ACE and Wind and 73P fragments, with the largest fragment B fixed as the origin and the direction of the Sun fixed along the Z axis. The blue regions shown in Fig. \ref{fig:3dax} provide a rough approximation of the extent of the comae of the tracked fragments, with a radius of 5 million km. The true extent is certainly much less than 5 million km, this value is larger to account for possible non-tracked fragments that have radially deviated from the primary trail. Plotted with the trajectories is the region swept by the hypothesised debris field, trailing behind fragment 73P-B. To determine this region, the average orbit vector of the known fragments was calculated. In the rotating frame as plotted in Fig. \ref{fig:3dax}, this vector traces the shaded area. In reality, the dust trail deviates from and encloses this vector, so the shaded regions should be treated as estimates of the path of the dust trail, as a visual guide. Plotting in this frame of reference limits the motion of the two spacecraft to be relative to the cometary fragments. In the upper panels of Fig. \ref{fig:3dax}, the radial solar wind flow is oriented left to right, while in the bottom panel, the radial wind flows out of the page.

These plots demonstrate the motion of the cometary fragments relative to the spacecraft at SEL1, providing an insightful view of the encounter. It can be clearly seen in each panel that the O$^{+}$ flux peaks align well with the estimated trail region, rather than the primary fragments. This is particularly apparent in the upper panel,  where the Sun is to the left, and the lower panel, which is oriented towards the Sun. The cometary fragments are widely distributed along the original orbit of 73P, but are more tightly distributed perpendicular to the average orbital velocity vector.


\section{Observations - STEREO-B}


\begin{figure}[ht!]
\centering\includegraphics[width=1.\columnwidth]{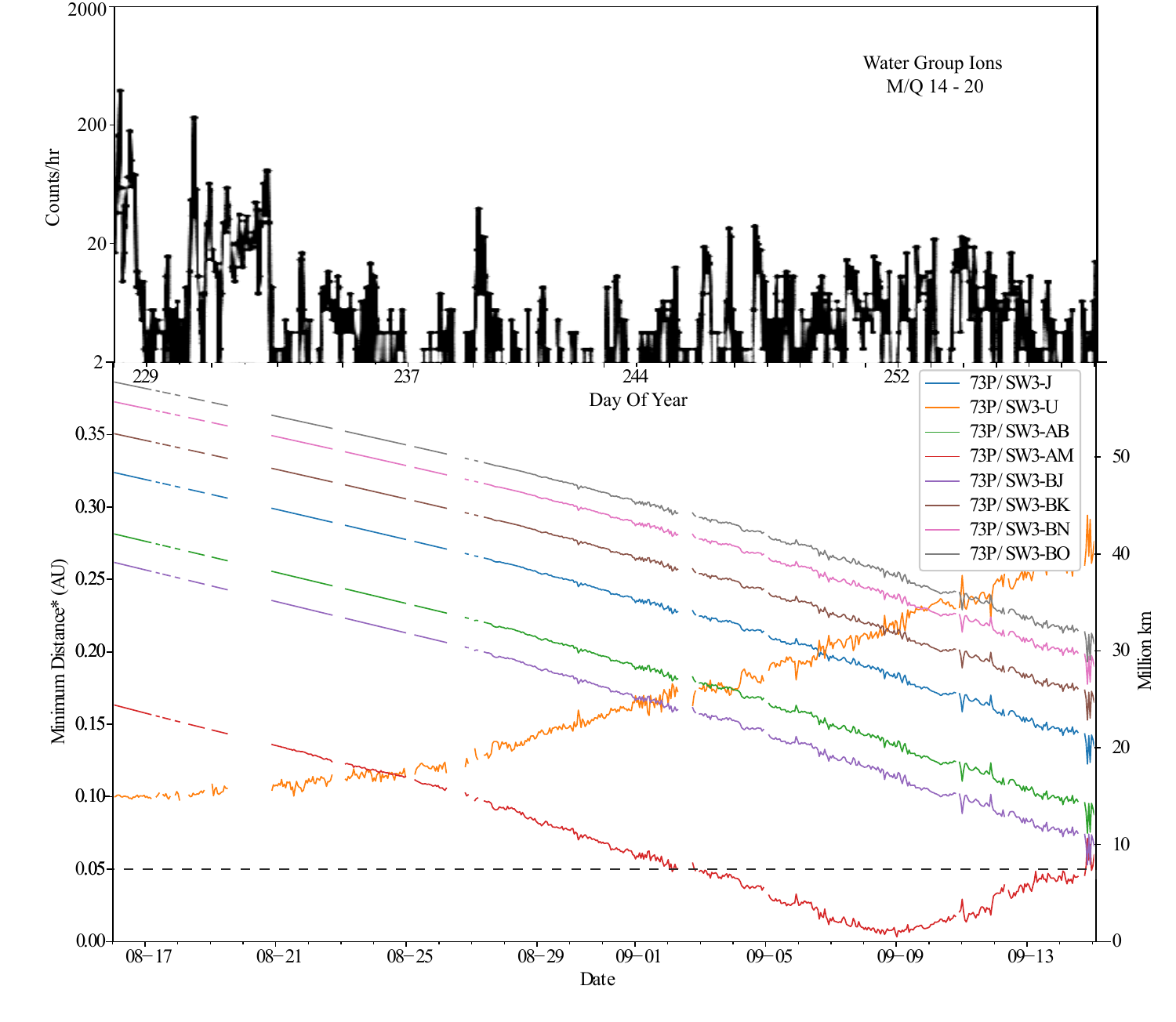}
\caption{\textit{Top:} Ion count rates recorded by STEREO-B/PLASTIC in the mass-per-charge range of water group ions (14-20). Ion fluxes are not available during this time period. \\
\textit{Bottom:} Impact parameters calculated for 3-dimensional solar wind velocities measured by STEREO-B, for fragments of comet 73P. }
 \label{fig:stb1}
\end{figure}

In August 2011, fragments of comet 73P passed directly sunward of spacecraft STEREO-B, during which there was an increased flux in water-group cometary ions detected by STEREO-B/PLASTIC (PLAsma and SupraThermal Ion Composition), coincident with comet C/2010 X1 (Elenin) passing sunward of STEREO-B \citep{galvin2015}. Using proton velocities from PLASTIC data to calculate the impact parameters for the fragments of comet 73P, a previously unknown ion tail crossing of comet 73P fragment AM is identified. This is shown in Fig. \ref{fig:stb1}, alongside the ion count data from STEREO-B/PLASTIC, where a subtle peak is visible, approximately aligning with the impact parameter minimum for comet 73P fragment AM. 


When the comet reaches the orbit of spacecraft STEREO-B, the impact parameter is calculated as being below 0.05AU, or $7.40{\times}10^{6}$ km, considered to be a suitable threshold for cometary ions being encountered and accounting for some assumed ion tail broadening. This coincides with an abrupt flux increase in water-group ions detected by STEREO-B/PLASTIC. The impact parameter remains this low for the duration of the flux peak, before increasing as the comet continues to approach the Sun. Given that the predictive method used to calculate the impact parameters produced results that are in agreement with in-situ measurements, an alignment, similar to that observed with comet Elenin, appears to have occurred with fragment 73P-AM. The resulting cometary ion flux from this encounter is dependent on the level of ion production at the fragment and the resultant ion number density in the tail. Fragment 73P-AM is significantly smaller and less active than the larger fragments present in the years preceding, so a smaller flux is expected. Due to the limitations of the PLASTIC instrument, the actual presence and cause of this water-group ion flux peak is tentative, but the alignment of the spacecraft with the projected ion tail of the fragment upstream indicates that some small portion of ion flux did result from the fragment 73P-AM.

\section{Discussion of results}

As discussed previously, the WIND/ACE event of 2006 was originally identified by \citet{gilbert2015}. In addition to the unusually high and sustained O$^{+}$ ion flux, elevated proton densities, and enhanced OH$^{+}$/H$_{2}$O$^{+}$ concentrations were also identified. This most clearly occurs 4 times over the weeks of the encounter \citep{gilbert2015}. These various signatures all point to a cometary source, and given its proximity to the spacecraft during detection, 73P is overwhelmingly the most likely candidate source. However, Figs. \ref{fig:3dax} and \ref{fig:impa2} discount the primary fragments of 73P as sources. This discrepancy between the fragment orbital plane crossings and the flux peaks is attributable to the presence of non-tracked sub-fragments in the trail. The extended dust trail of 73P is particularly prominent, being visible in Fig. \ref{fig:dust_trail_example} from May 2006 \citep{spitzer}. 73P had undergone further fragmentation in the weeks before the 2006 encounter, exposing trapped volatiles to the Sun. Sub-fragments making up the dust trail, too small to be individually identifiable, could quickly become depleted. Nonetheless, the combined outgassing from numerous freshly exposed sub-fragments could be enough for the extended trail to generate a significant ion stream or sheet. Such an ion sheet would account for the significant length of the encounter and the relatively sustained enhancement in the O$^{+}$ flux.

The most significant inaccuracy of the Tailcatcher method may be the approximation that the solar wind velocity does not change with radial distance from the source object. A Parker model assuming a purely radial velocity would provide an approximation of the evolution of the radial flow, but this would still be significantly inaccurate if some additional object was present between the spacecraft and the comet. For shorter separations, the deviation resulting from the constant velocity assumption is reduced. Both the 2006 and 2011 encounters occurred at very short spacecraft-comet distances. Additionally, even if the Tailcatcher results in Figs. \ref{fig:stb1} and \ref{fig:impa2} are discounted, the alignment of the encounters is confirmed visually by inspection of Figs. \ref{fig:3dstb} and \ref{fig:3dax}.

For the 2011 STEREO-B encounter, while there are likely some weak enhancements in ion fluxes at the spacecraft resulting from the trail of 73P, the source of water group ion flux is likely to be the fragment closest to alignment, 73P-AM. In contrast to the 2006 ACE/Wind encounter with 73P, this later encounter demonstrates what is expected from an encounter with the ion tail of a significant cometary fragment. There is close agreement between the minimum impact parameter and ion flux peak, and the minimum impact parameter reduces to well below 0.05AU. This further demonstrates the poor alignment in the 2006 encounter, where the fragments' impact parameter minima neither get close to 0.05AU nor occur concurrently with the O$^{+}$ flux peaks measured by the two spacecraft. The prediction method used in both cases is identical, and there does not appear to be another known comet somewhere sunward of SEL1 that could be the O$^{+}$ source.

Given the apparent confirmation of detectable ion flux originating from certain cometary dust trails like 73P, the Tailcatcher method could now be adapted to search for past and future serendipitous trail crossings. Similar encounters may have gone unnoticed in older spacecraft data-sets due to the low prominence of any signatures, but could be identifiable using this method. The methods used in examining these encounters with 73P could be applied to search for cometary dust trails upstream of detected interplanetary magnetic field enhancements. This would involve assuming the presence of putative dust trails for various comets and asteroids, even if not directly observed.

The mechanics of the production of pickup ions from small sub-fragments are not explored here, but would be better investigated with similar encounters to compare. These observations may have implications for trails of material that are deposited in the paths of asteroids or extinct comets, which may have similar observable signatures, albeit likely far weaker due to their expected rapid depletion of volatiles caused by a larger surface to mass ratio than a nucleus would have. Though not analysed here, 73P encounter with ACE and Wind provides a potential measurement of the kind of magnetic interference and draping that could be present from the dust/fragment trail, at near distances. These magnetic features would also be present at greater distances from the source, although diluted and weaker. Modern spacecraft with particularly sensitive magnetometers and ion spectrometers may be capable of measuring cometary pickup ions and the influence of cometary debris trails on the HMF at greater distances from the source. However, encounters with similar ion sheets that are significant enough to be measurable appear to be very rare.

\section{Conclusions}

Recently fragmented comets such as 73P provide opportunities to study their chemically pristine interior composition, and the transport of cometary ions by the solar wind allows measurements to be made millions of kilometres from the fragments.
The 2006 encounter in particular presents a unique case that is only detectable due to significant ion production by the extended debris field along the orbit of 73P, resulting in an ion ``sheet'' lagging behind the primary tails of the largest leading fragments.
The cometary ions encountered by the spacecraft Wind and ACE located at SEL1 have their source further behind the known tracked fragments of comet 73P, in the extended debris trail. It is unlikely that the majority of ions detected are sourced in the extended coma surrounding the larger fragments, as ion flux continues to peak at spacecraft-comet separations much greater than the expected extent of the coma. 
Cometary fragments that are too small to be detectable visually can nonetheless be detectable sources of cometary mass loading to the solar wind, and may lead to magnetic fluctuations that persist far downstream of the nucleus.
These encounters provide a model for the Tailcatcher program to be expanded to identify similar encounters where the magnetic interference and pickup ions resulting from the trail could be encountered.


\begin{acknowledgments}
\section*{Acknowledgments}
We thank the STEREO-B/PLASTIC, ACE/SWICS, and Wind/STICS instrument teams for the use of data in this work. We are grateful for the indispensable work of J. A. Gilbert and co-authors for their identification and analysis of the 2006 in-situ encounter of 73P/SW-3 with spacecraft ACE and Wind. 
Author SRG acknowledges the support of a UKRI Science and Technology Facilities Council (STFC) research studentship. Author GHJ acknowledges support by STFC grants ST/S000240/1 and ST/W001004/1.

\end{acknowledgments}

%

\vspace{5mm}
\newpage




\newpage
\bibliography{refs2}{}
\bibliographystyle{aasjournal}



\end{document}